\documentclass{article}
\usepackage[utf8]{inputenc}
\usepackage{ismir,amsmath,cite,url,graphicx}

\title{Polyphonic pitch detection with convolutional recurrent neural networks}

\twoauthors
  {Carl Thomé} {DoReMIR Music Research AB \\ {\tt carl.thome@doremir.com}}
  {Sven Ahlbäck} {DoReMIR Music Research AB \\ {\tt sven.ahlback@doremir.com}}

\def\authorname{Carl Thomé, Sven Ahlbäck}

\begin{document}

\maketitle

\begin{abstract}
Recent directions in automatic speech recognition (ASR) research have shown that applying deep learning models from image recognition challenges in computer vision is beneficial. As automatic music transcription (AMT) is superficially similar to ASR, in the sense that methods often rely on transforming spectrograms to symbolic sequences of events (e.g. words or notes), deep learning should benefit AMT as well. In this work, we outline an online polyphonic pitch detection system that streams audio to MIDI by ConvLSTMs. Our system achieves state-of-the-art results on the 2007 MIREX multi-F0 development set, with an F-measure of 83\% on the bassoon, clarinet, flute, horn and oboe ensemble recording without requiring any musical language modelling or assumptions of instrument timbre.
\end{abstract}

\section{Introduction}
Deep learning has dramatically improved the state-of-the-art in speech recognition\cite{lecun2015deep} and therefore it seems likely to be a promising framework for music information retrieval (MIR) tasks like automatic music transcription (AMT), as AMT shares many similarities with ASR in the sense that methods often rely on estimating sequences of events from input sequences with fixed sampling rates such as spectrograms (\figref{fig:cqt}).

\begin{figure}
\includegraphics[width=\columnwidth]{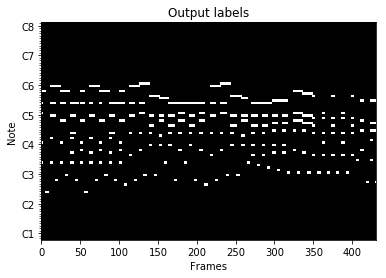}
\caption{Example of an output piano roll with MIDI notes drawn over time frames.}
\label{fig:piano_roll}
\end{figure}

\cite{kelz2016potential} showed that deep neural networks can learn to transform spectrograms to piano rolls (\figref{fig:piano_roll}) with no explicit assumptions about musical structure or instrument timbre, and particularly convolutional layers seem beneficial to avoid overfitting. \cite{troxelmusic} used skip-connected convolutional networks as described in \cite{he2016deep} to guess active fundamentals around predetermined onsets (and then joined notes by always extending offsets to be performed as legato) with an average onset-only F-measure of 83\% on the MIREX 2016 solo piano note tracking task.

However, as music is sequential it should be beneficial to promote temporal dependencies in the model architecture and previous studies have used recurrent neural networks (RNNs) to do so \cite{bock2012polyphonic} with average onset-only F-measures of 92\% for piano recordings. However, a problem with RNNs is that the recurrence is fully-connected \cite{sigtia2016end}, meaning deeper networks will overfit, and essentially memorize timbre and/or melodies as spatial structure is discarded. To circumvent overfitting \cite{bock2012polyphonic} used semitone filterbanks, but we believe that distinguishing pitch in polyphonic signals requires a high frequency resolution such that spectral peaks don't overlap. Essentially the spectrogram has to be invertible for us to believe that a model gets a fair shot at learning.

This work's main goal is to propose a combination of convolutional neural networks and recurrent neural networks to perform polyphonic pitch detection without requiring musical language modelling or assumptions of instrument timbre. This approach is inspired by the nature of human pitch perception, which occurs on a sensory level and can be regarded fundamentally a psychoacoustical phenomenon. 

\section{Related work}
Many variants of combining convolutional neural networks with recurrent neural networks have been tried \cite{amodei2015deep, choi2016convolutional}, often by stacking networks such that the output of a convolutional model is directly fed into a recurrent model, and trained jointly. This is straightforward but still discards spatial structure in the recurrent part of the network. Instead, \cite{xingjian2015convolutional} proposed replacing the recurrent connection with convolutions (ConvLSTM), and applied a seq2seq \cite{sutskever2014sequence} variant to precipitation nowcasting\footnote {Loosely speaking, precipitation nowcasting is the problem of predicting if it will rain from satelite images.}. ConvLSTMs were later applied to spectrograms in ASR \cite{zhang2016very}, although they pool over time before the RNN as their output is only a few symbols per sample (words in sentences) compared to the output resolution of piano rolls (hundreds of frames in order to put notes at accurate onsets). Despite needing high time resolution, we choose to work with piano rolls to avoid having an encoder and decoder, and to benefit from forcing models to choose a MIDI note number regardless of tuning and inharmonicity. Future work will look into alternative output representations, like attention seq2seq models where the output is a list of pitch events instead of piano rolls.

Aditionally, \cite{zhang2016very} used rectified linear units (ReLUs) \cite{nair2010rectified} and batch normalization \cite{ioffe2015batch} which means they might have been hindered by trouble described in \cite{cooijmans2016recurrent}. They settled on only normalizing input connections and not the recurrence, as proposed by \cite{amodei2015deep}. For seq2seq this might not pose a problem, but for polyphonic pitch detection it's important to normalize feature maps over time because the output is essentially a direct translation of the input with a recoloring of the spectrogram into a binary piano roll.

\section{Problem formulation}
MIREX makes a distinction between estimating active fundamental frequencies with fixed sampling rate, called "framewise evaluation", and estimating note events with determined start and stop times, called "note tracking". Both tasks are evaluated separately. In this submission we're concerned with determining note events on the MIDI scale, instead of tracking pitch contours, although the latter will be investigated in future work for vibrato tracking, note-level instrument recognition and overlapping polyphonicity in ensemble recordings (not representable in piano rolls).

For both note tracking and framewise multi-f0 estimation, two things make the problem hard:
\begin{enumerate}
\item When multiple notes are played together all tones are mixed together in the same audio signal, so there is more noise and it can be hard to distinguish fundamentals from partials, particularly when partials overlap to create false fundamentals.
\item Human perception of pitch is complicated. Pitch often corresponds to the fundamental frequency of a sound wave, or the presence of a harmonic series, but not always. Timbre is also important for pitch perception \cite{oxenham2012pitch}.
\end{enumerate}
In other words, for AMT it can be better to have a pitch detector that adds non-existing frequencies, because humans generally perceive pitch categorically as stable note events, e.g. we hear notes in melodies and unconsciously modify the sound to fit musical expectations.

\section{Method}

\subsection{Preprocessing}
\begin{figure}
\includegraphics[width=\linewidth]{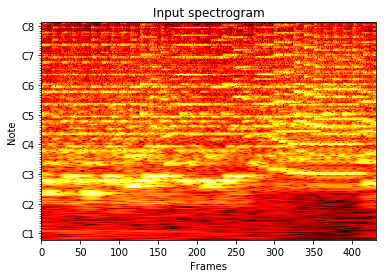}
\caption{Example of an input spectrogram where the audio has been constant-Q transformed. Note that only log-scaled power is shown as predicting phase is still an open question in MIR research.}
\label{fig:cqt}
\end{figure}

We use the real-time variant of a non-stationary Gabor transform where Q is constant called sliCQ\cite{holighaus2013framework} (\figref{fig:cqt}), but base our implementation on the time-domain implementation of the constant-Q transform (CQT) in \cite{mcfee2015librosa}. The reasoning for relying on CQT is that it promotes pitch invariance compared to an ordinary short-time Fourier transform (STFT)\cite{benetos2012shift}, which is important for convolutional neural networks that share weights over the frequency axis. As the presence of partials is also important for pitch perception we stack multiple spectrograms in the depth dimension as in \cite{bittner2017deep}, although we only use harmonics 1, 2, 3 and 4. All spectrograms are log-scaled power spectrograms so convolution kernels can suppress noise as well as detecting harmonics (which are often a lot louder).

\subsection{Model}
The model is a deep neural network that transforms spectrograms (\figref{fig:cqt}) to piano rolls (\figref{fig:piano_roll}). We treat the occurrence of a new note onset as orthogonal to the presence of pitch, so we have two separate output channels, one for articulation and one for sustain (\secref{sec:postprocessing}). The sustain channel is zero everywhere except during note events where it is set to ones (i.e. a typical piano roll representation). The articulation channel has the additional constraint that it is only one at the start of a note event (i.e. the first-order derivative of the sustain channel, without negative values).

The network performs a succession of convolution and average pooling operations in frequency to remove inharmonicity and timbre. The following recurrent layers have fully-connected matrix multiplies replaced with convolutions to preserve spatial structure in each feature map. We then upsample over time and apply additional convolutions to account for spectrogram frame overlap. Scaled exponential linear units are used as activation functions and scaled hyperbolic tangents are used in the recurrence \cite{clevert2015fast, klambauer2017self}. Layer normalization\cite{ba2016layer} is used over time, which unlike \cite{cooijmans2016recurrent} has the benefit of not requiring too many parameters. As gating we use long short-term memory cells\cite{hochreiter1997long}. We have also experimented with exchanging the ConvLSTM with a ConvGRU. The GRU \cite{cho2014learning} has been shown in \cite{greff2016lstm} to work as effectively as the LSTM \cite{hochreiter1997long} but it is our belief that uncoupled forget gates are important for mapping spectrograms to piano rolls as songs consist of different sections that can change suddenly, so we keep all LSTM gates with uncoupled forget gates\cite{gers1999learning}, and also add peepholes\cite{gers2002learning} for the same reason. Skip-connections\cite{he2016deep} are added around each recurrent, pooling and upsampling layer to encourage later layers to make adjustments early in the training, and were preferred over highway networks because gradient flow is improved without extra parameters. We also believe that the residual style is particularly suited for suppressing overtones. The final layer is a network-in-network (i.e. 1x1 convolution) with sigmoid outputs as in \cite{kelz2016potential, troxelmusic}.

Training was done with Adam\cite{kingma2014adam} and truncated backpropagation through time with RNN state passthrough across minibatches. Parameters were initialized with orthogonal initialization\cite{saxe2013exact}.

\subsection{Postprocessing}
\label{sec:postprocessing}
Finally, we peak pick the two-channel activation matrix to convert the framewise piano roll to a list of note events. Per note, we step through each time frame and place an onset at positions where the articulation channel is above a set threshold, and then include all frames onward until the sustain channel is under another fixed threshold, at which point we output an offset. If a new articulation is found during an active note event we simply fragment it by outputting additional offsets and onsets.

\section{Results}
To be announced as MIREX 2017 is reported.


\section{Conclusions and Future Work}
Our polyphonic pitch detection system gets state-of-the-art results on the 2007 MIREX multi-F0 development set, with an F-measure of 83\% on the bassoon, clarinet, flute, horn and oboe ensemble recording without requiring any musical language modelling or assumptions of instrument timbre.

A weakness of our approach is that if we want to extend our method with vibrato detection or ADSR envelopes, we would need to produce such annotations first. Therefore, for future work we are looking into complementing the deep learning approach with traditional signal processing techniques such as partial tracking\cite{mcaulay1986speech} although we strongly believe in a data-driven, end-to-end approach that avoids hard-coded heuristics.

Still, the primary weakness of our approach is the rigidity of piano rolls. There's no natural support for same-note polyphonicity and piano rolls contain a lot of redundancy. If we were able to supplant them with note events directly (e.g. seq2seq) that would go a long way \cite{zhang2016very} but it is not straightforward for AMT as there are invariances not covered in an events list because of the polyphonicity. For example, it shouldn't matter which order simultaneous note onsets are listed in (e.g. a chord). Data augmentation to represent invariances should be tried, but might inflate training times. Ultimately, we believe the major bottleneck for further progress is the piano roll representation.

\bibliography{references}
\end{document}